# The value chain of Industrial IoT and its reference framework for digitalization

(工业物联网的价值链和参考模型研究)
Hang song[1*], Yuncheng jiang[2]

Nowadays, we are rapidly moving beyond bespoke detailed solutions tailored for very specific problems, and we already build upon reusable and more general purpose infrastructures and tools, referring to them as IoT, Industrial IoT/Industry 4.0[1-3], etc. These are what will be discussed in this paper. When Industrial IoT (IIoT) is concerned about, the enormous innovation potential of IoT technologies are not only in the production of physical devices, but also in all activities performed by manufacturing industries, both in the pre-production (ideation, design, prototyping) and in the post-production (sales, training, maintenance, recycling) phases.

It is also known that IIoT acquire and analyze data from connected devices, Cyber-Physical Systems (CPS), locations and people (e.g. operator); along with its contemporary new terms, such as 5G, Edge computing, and other ICT technologies with their applications[4]. More or less it is drawn upon on its combination with relative monitoring devices and actuators from operational technology (OT). IIoT helps regulate and monitor industrial systems [2], and it integrates/re-organize production resources flexibly, **enhanced OT capability in the** smart value chains enabling distributed decision-making of production.

## 1. Introduction

Towards the IoE (Internet of Everything) era, the traditional, fragmented processes of design, production and customer fulfilment will be replaced by the management of the end-to-end design-to-customer fulfil, where products are designed based on customer requirements (granularity decided by customer-focused manufacturing) and their producing cycles are shorter. And, the processes do not finish with product delivery; the product-service provides information for the maintenance services and for continuous design of products and processes. Autonomous sensors in machine and manufacturing services will facilitate the operational performance model for predictive maintenance. Furthermore, three main reasons of IIoT value chain analysis are as followed.

a. Autonomy and the internal self-balance mechanism in manufacturing are needed in a closed loop of IoT enabling configuration dynamically and monitoring of the operational capabilities of the company. So IIoT concentrates on networks in/with plants, quality control and efficiency improvement. But nowadays it is still on its beginning. The autonomy production and beyond, e.g. in the close-loop of production-consuming-usage-upgrading/ elimination-reproduction, information dual-directions are needed in their self-balance, i.e., after the interfaces and data sets are given, the smart agent for manufacturing beside when to work, how to distribute the work force(e.g. flexible manufacturing), with the automation, as well as the combination of Machine intelligence (e.g., Machine Learning).

b. **Enhanced OT capability** integrated **in the** smart value chains enabling distributed

decision-making of production. The promise of the IIoT includes significant **opportunities** in the coming years. Management consulting firm, McKinsey & Co., estimates that the IIoT will create $7.5T in value by 2025. The Industrial IoT brings together minds and machines—connecting people to machine data that accelerate digital industrial transformation[1].

**c.** Flexible integrated of production resources (workforce, information, sensors and networking) makes the autonomy of IoT across the grid (resources) easier.Figure 1-1 and Figure 1-2 show modular machine design with HM collaboration and machine adjustment with Sick's appropriate technology in man-made wood procession[2] respectly. And Figure 1-3 shows the Continuous condition monitoring that can be carried out in the servo for actuator and remote diagnose on-line.

In short, Industrial Internet of Things (IIoT)connecting business and processes – across the grid borders of heterogeneity，resource reallocation and service from companies, plant the OT(real-time supported or not) into different plants, embracing ICT in a super intergraded supervised scenario and pushing the service, security, open and intelligence into smart manufacturing of new stage.

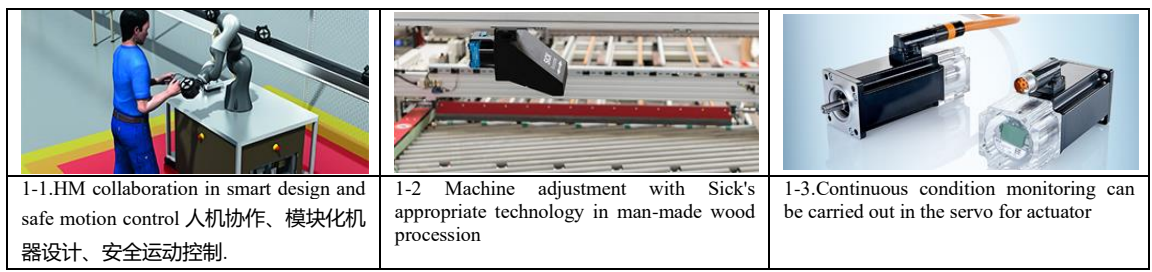

| 1-1.HM collaboration in smart design and safe motion control 人机协作、模块化机器设计、安全运动控制. | 1-2 Machine adjustment with Sick's appropriate technology in man-made wood procession | 1-3.Continuous condition monitoring can be carried out in the servo for actuator |

Figure 1 Flexible integrated of production resources (From: "https://www.sick.com/")

## 2. Industrial IoT and I4.0

Compared to IoT, Industry 4.0(I4.0), Smart Manufacturing and IIoT are relatively new terms, whose understanding depending on the source and their context, and their descriptions or usage of these terms may vary significantly. Thus, it is important to discuss their specific definitions and clarify the understanding of them.

IIoT is defined as IoT applied in the industrial environment, by ENISA (European Union Agency For Network and Information Security) in the Baseline IoT Security Recommendations[3]. Industry 4.0 is, in turn, a much broader concept that encompasses IIoT and Smart Manufacturing/Factory alike, in the fourth industrial revolution era.

As for IoT, whose ultimate goal of the development is to realize the complete integration of human society, information world and physical world; is on its long term evolution way towards IoE (Internet of Everything). Industry 4.0 synchronizes with this goal, and tries to build a controllable, credible, scalable, safe and efficient physical device interaction in the industrial scene, and try to fundamentally change the traditional human understanding of industry, as shown in Figure 2.

---

[1] https://www.ge.com/digital/blog/what-edge-computing#to-section-index=section-3
[2] http://www.sickinsight-online.de/diagnosedaten-frei-haus/
[3] See ENISA (2017) "Baseline Security Recommendations for IoT":
https://www.enisa.europa.eu/publications/baseline-security-recommendations-for-iot/at_download/fullReport

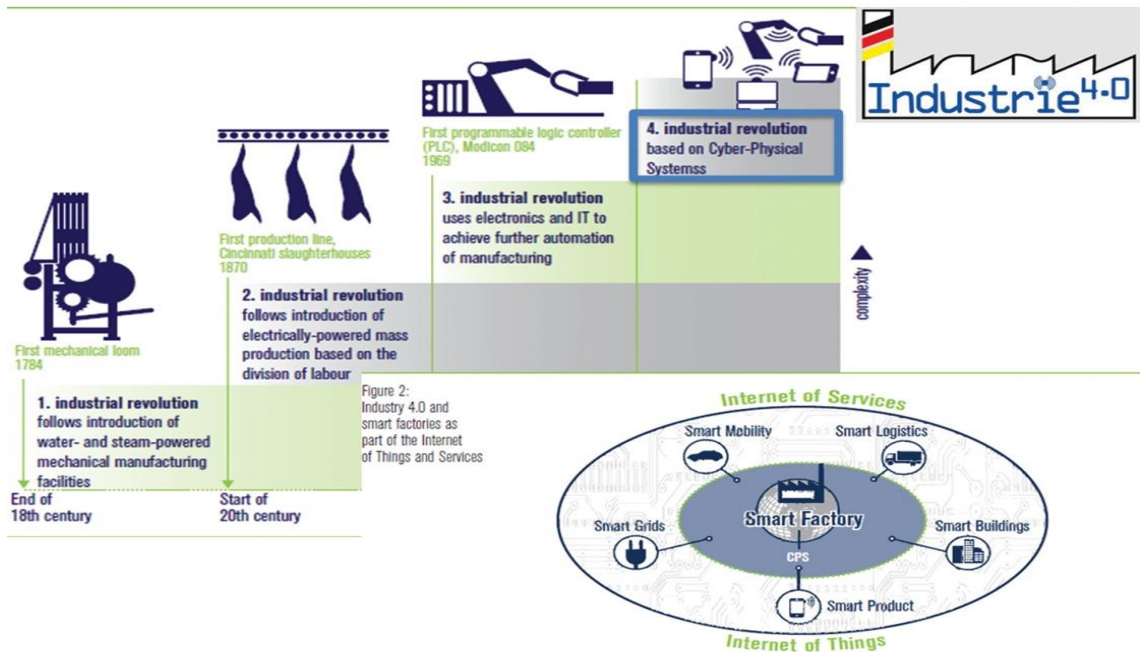

Figure 2 Steps from 1st Industrial revolution to the 4[th] Industrial revolution [2]

In the upgrading steps, the traditional industrial barriers will disappear, and a variety of new activities and cooperation forms will be produced in the practice of the innovation economics.

**2.1. I 4.0**

Nowadays, the role of the IIoT is becoming more prominent in enabling access to devices and machines, which in manufacturing systems, were hidden in well-designed pattern, principle and methodology of Industry 4.0. Industry 4.0 is just a pronoun for the new fourth industrial revolution, including interoperability, autonomy, information transparency, technical assistance and distributed decisions[4], which makes the traditional manufacturing smart (see the right corner of figure 2).

The concept of I4.0 includes the basic mode transformation from centralized control to decentralized enhanced control, and the goal of "shift" is to establish a highly intelligent and digital industrial production and organization mode. The shift is emphasized by ENISA on its Industry 4.0 definition, as "a paradigm shift towards digitalized, integrated and smart value chains enabling distributed decision-making in production by incorporating new cyber-physical technologies such as IoT"[5].

In the "shift" of applying new generation information and communication technology (ICT) to create new value chains, the division of labor and resource in industrial chain will be reorganized optimally and efficiently. The value chain of industry 4.0 has three supporting points: firstly, product leads production and ICT technology services the production. The second is to make the machines/devices more intelligent, able to talk to each other in their collaboration and even communicate with products and production (before and afterward, e.g. designing, prototype, planning and maintaining). The third is to make the products reflect the social value of environmental protection, safety, reliability and sustainable development.

---

[4] See Connected Factory Global (2016) "Manufacturing Control System Cybersecurity: Risk Assessment & Mitigation Strategies": http://www.connectedfactoryglobal.com/resources/cybersecurity-report/

IIoT-wise, Industry 4.0 is an innovative " sets of rules and connection " of industrial manufacturing and resource distributing in vertical application of **IoT grids[5] and IoT zones**（**see Figure 3**）.

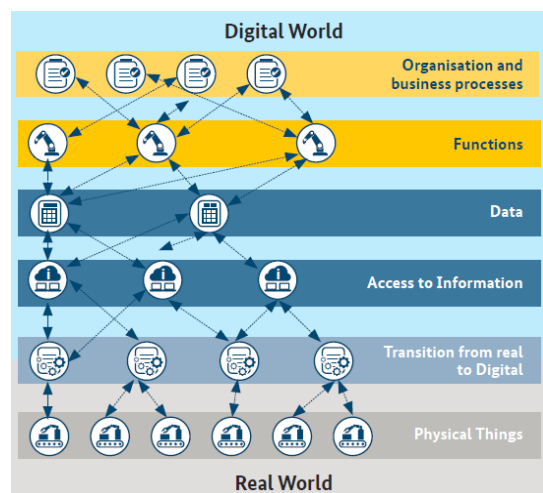

Figure 3. IoT Zones and conduits in I4.0, where the communication/connection can be organized either in 1-to-1 or m-to-n relations as required by the related business functionalities [6].

Recent progress that resulted in reversal of the traditional production process logic led to the formation of this concept, representing a shift towards decentralized production. According to the shift and its companying ICT enabler, industry 4.0 is divided into Zones and conduits with a layered architecture in Fig.3. Zones can be divided into grids and even subdivided into sub-grids (e.g. Primitive IoT[6]) according to different manufacturing scenarios. Under the same rules, the sub-division of vertical IIoT in Industry 4.0 integrates with each other and develops **collaboratively with its** communication relationships (in Fig.3). The value of intelligent manufacturing itself is multi-objective, not only to do a good job in one product or a single production line, but also to minimize the waste in the production process widespreadly.

Firstly, the IIoT should aim at the cost performance ratio of the whole life cycle of the product; the design and manufacturing process, including post maintenance and reproduction, can be matched with the user demand and the ideal price (forecasting, evaluation and other statistics of data science). The **second** is to make the whole system smart enough for self-awareness automatically adjustment according to the change of the product condition (e.g. demanding/requirement) in the manufacturing process, and realize the real-time "autonomy (or self-introspection)" of the system status based on the dynamic planning .

The **third** is to realize zero failure, zero hidden danger, zero accident and zero pollution in the whole manufacturing process, which is the highest level of smart manufacturing in future IIoT (maybe realized in Industry 5.0 or beyond, and the authors just say so here).

The infrastructure design and the layout of zones and conduits in such solutions need to be developed in line with the results of the security risk assessment, which has to consider the possible impact on the safety risk reduction required by the I4.0 and IM (Intelligence Manufactory) workspace.

In a long run, either the shift of intelligent/ digital in production and organization, or the shift of decentralized production, we are just on the way to intelligent IIoT. So the value chain we desired for IIoT exists as the corresponding three points: 1). through industrial automation / industry 4.0/industrial Internet and IIoT, the horizontal integration of resources between (within) enterprises / factories can be realized. 2). Integration and intelligent application of end-to-end engineering digitization throughout the whole value chain. 3). within the enterprise (factory), vertical integrated and networked manufacturing system (beyond the horizontal combination/re-organization) can be flexibly.

---
[5] IoT grid is presented by the first author of this paper in reference [1].
[6] Service oriented Primitive IoT is presented by the first author of this paper in reference [1].

1.通过工业自动化/工业 4.0/工业互联网，以及网络，实现企业间/企业内（工厂内部）的资源横向集成。2.贯穿整个价值链的端到端工程数字化的集成与应用。3.（I4.0 时代）在企业（工厂）内部，实现灵活地，可以重新组合的纵向集成和网络化的制造系统

As Figure 4 shown, at the beginning stage of intelligence as Device-centric IoT, a product/device is not merely processed by machines – it communicates with other devices in producing context where providing relevant information and instructions . When production related device in manufacturing (e.g. production lines) is no longer isolated, and devices of Device-centric IoT have become integral parts of the overall network by communication and distribute computing (e.g. Edge or cloud computing, especially in discrete manufacturing), the collaborative-intelligence stage upgrades into the System-centric IoT stage, which introduced complex analytic and flexible decentralized systems in the Networked intelligence environment.

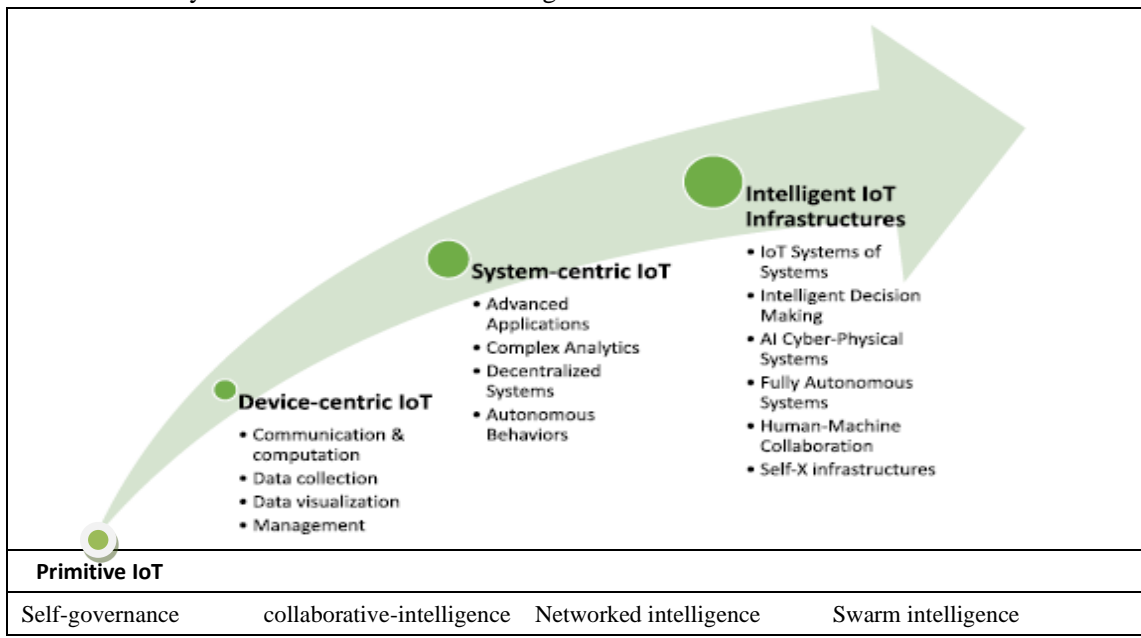

Figure 4 Service oriented Primitive IoT and different stages towards Intelligent IoT (Adapted from:[4])

The character of System-centric IoT is that its functions are not bound to hardware but distributed throughout the network. In these new systems internal communication can now be observed across an organization's hierarchical levels, which will be introduces in Fig.5, Fig. 6 and Fig.7.

System-centric IoT connects production to information and communication technologies. It merges end user data with machine data and enables machines to communicate with each other. As a result, it has become possible for components and machines to manage production autonomously in a flexible, efficient and resource-saving manner.

Intelligence –wise, new characteristics have been shown in **Intelligence IoT Infrastructures** (see

Figure 4), which including not only the IoT systems of systems, but also the AI ability both on different CPSs and flexible organization of external interactions.

**Intelligence –IoT Infrastructures** benefits include, among others, higher product quality, greater flexibility, shorter product launch times, new services and business models. It is important to note that the data flows depicted in the decision making or autonomous systems, refer to physical productions and data (e.g. exchange of digital twins). Data flows are at least bidirectional, e.g. customers may provide feedback to the production/manufacturing process [4].

**2. 2 Vertical IIoT Grids**

At the recent international symposium, a rapid survey on the application and value creation industry of the Internet of things shows that the application of the Internet of things in the industrial field will benefit the most[1]. The survey selected six different perspectives with "Internet of things, value and industry" as 1) IIoT applications now vs.5 years, 2) Tech./development impacting IIoT applications, 3) Obstacles or challenges, 4/5)Value from IoT systems-ACTIVE /PASSIVE and 6) your vision. The result analysis is shown in Figure 5.

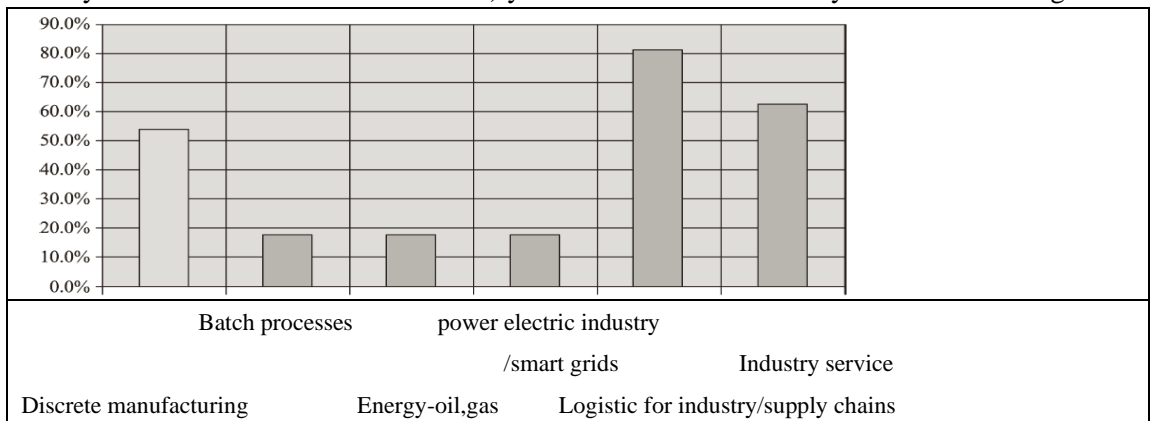

Figure 5 The sub-division industry of IIoT

The top three benefits of IIoT applications are logistics for industry and supply chains, discrete manufacturing, and industrial services. The rest are batch processes, energy (oil and gas), and power electric industry (smart grid).

However, the service improvement and production discretization in intelligent manufacturing are still urgent problems. The application of Internet of things in industry 4.0 is not only reflected in the customized solutions, but also in different IoT vertical Industrial services, where the IoT grid seems to be more suitable for the integration, deployment, configuration and usage(see Fig.6).

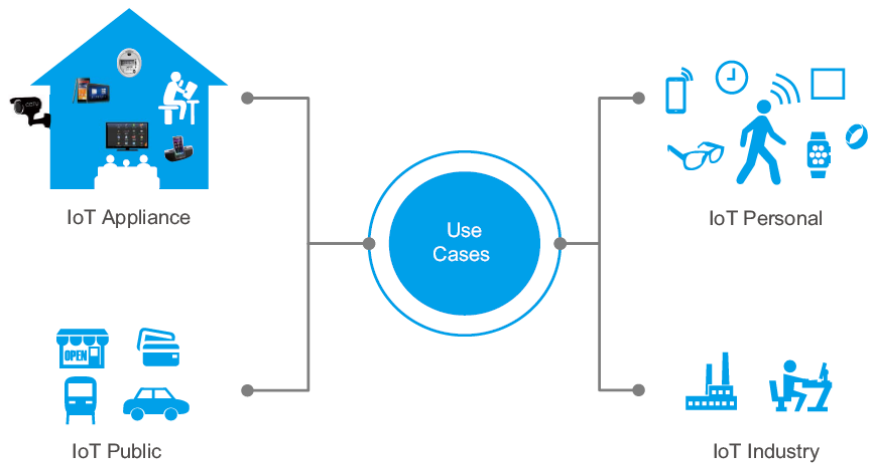

Figure 6 Four parts (IoT grids) as use cases grids/examples of IoT vertical Industrial services

In Fig.6, from the perspective of industry, IoT industrial is no more than a vertical grid of industrial Internet of things. The core reason is that it needs not only the installation and configuration of equipment, but also the development direction of independent intelligence. In short, according to the concept of IoT grid, the "big grid" of industry 4.0 is divided horizontally into the Internet of things, cloud computing, big data management, intelligent equipment, etc. Vertically, it is subdivided into industrial logistics and supply chain, industrial production service, discrete industrial production, batch production and other fields.

From the application on the right side of Figure 5, IoT industry, as the industry Internet of things, is only an industry grid for the industrial application of the Internet of things. According to the industrial use cases of the Internet of things, there can also be personal grid (wearable grid), smart home grid, smart transportation grid, etc.

In succession, the concept of "Internet of things +" (industry) grid such as agriculture 4.0, business x.0, smart tourism x.0 will appear, and they show that the grid of various industries will sooner or later move to the independent intelligent road of vertical functional autonomy and horizontal information union. In the initial stage of this road, it is the specific application of the Internet of things grid in all walks of life. In the process of rapid industrial transformation and technological layout update by using the Internet of things technology, the IoT grid can lead to its intelligent way from inside. It is hoped that the IoT grid can provide the good thinking for the independent intelligent road of IoT, and more practical work of IoT grid needs further research.

The development of IIoT/IoT urgently needs standards. From the "horizontal perspective", each standard in the perception layer, network layer and (platform layer) application layer does not have enough open ability to provide cross platform integration ability and free flow of information in this layer; in the vertical view, the application specifications of each industry are still in the state of "separation" and lack of function (service) format.

However, the application of Internet of things is gradually moving from vertical and single solution to cross industry, cross application, cross standard collaborative application or multi-objective application. Therefore, there is an urgent need for a long-term **standardization** based on standards but higher than standards. In response to the background of digital economy, the internal standardization reconstruction of the "system of the Internet of things" is consistent with the self-renewal business model, product service portfolio, and customer multi-objective (such as security) requirements. The product concept of product-as-a-service can be embedded into the standardized "grid", and can develop in the direction of horizontal cross platform and vertical cross industry.

# 3. The value chain of IoT grids in 2D/3D views

**3.1 The value chain of two dimensional domain of grids**

The Internet of things Grid is not a standard, but a service-oriented regulation exploration or the Conceptual Model either links an entity in the physical world to the service it can afford (or enjoy) in order to be discussed detailly in an IoT industry (or certain business aim), or helps to decompose IIoT with a grid vision. The latter includes not only RAMI4.0, but Reference Architecture Model Edge Computing (RAMEC, by Fraunhofer FOKUS) **in Section Cloud and Edge computing, and other** Conceptual Model of Industry architectures in **Section IIoT & Manufacturing.**

In Figure 7, if we regard the horizontal X-axis of grid coordinates as the services of different industrial cases, and regard the vertical Y-axis as the layered different platforms, the hidden Z-axis will be explained as product view or product-as-a-service, which be clustered as M2x, B2x and G2x (Machine/Business to everything, Governments/Games to everything, which extended to wearable device/service). Along the X axis (as shown in the vertical direction of figure 7), vertical industries include agriculture, commerce; smart city, intelligent transportation, logistics, etc.

Along the Y axis, the horizontal layered platforms include perception and device/product layer, infrastructure and networks layer, data / software platform, and digital service layer.

The hidden Z- axis which is vertical to the x-y plane includes sensor as a service, systems of CPS as a service, wearable as a service, security as a service, product/production-as-a-service and so on.

**From the perspective of product, when IoT Grid is applied in the decomposed scenarios with product-as-a-service, the minimum granularity service of a product belongs to its corresponding PIoT.**

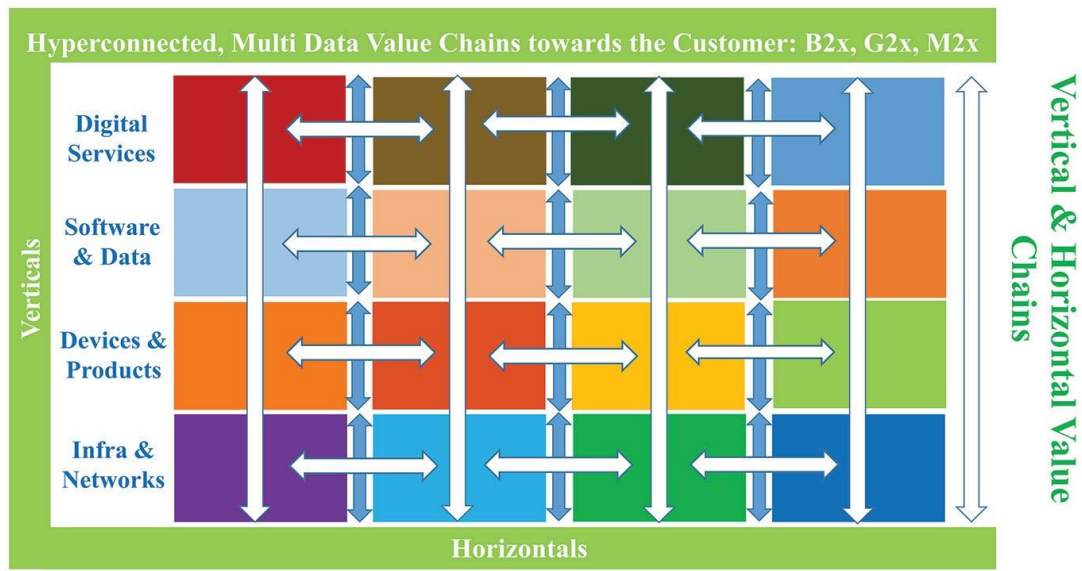

Figure7. Gridded Value chains of vertical and horizontal, where the hidden Z-axis is To-everything services (e.g. Business/ Games /Machine/production-as-a-service)

**3.2 A cubic-gridded analysis of RAMI4.0 and IMSAs**

The objects/devices / things, and places represented as Physical Entities are the same as **Assets** mentioned here, which are explained as "an item or property that is regarded as having value", according to the Oxford Dictionary. A Physical Entity can potentially contain other physical entities; for example, a building is made up of several floors, and each floor has several rooms. So IoT Zone/Domain is made up of several IoT Grids and each IoT Grid has several PIoT (e.g., parking grid includes: the car, the right parking location as a certain goal, how to part the car as the relaitions of the former two). Therefore, the term Asset is more related to the business aspects of IoT than objects/devices / things.

In some occasions, a Human User can choose to interact with the physical environment by means of a service or application, and the application is also a User when it needs service from other IoT Grids in the opinion of IoT Grids model.

In Figure 8, the RAMI4.0 [7] (Reference Architectural Model Industry 4.0) is a service-oriented technical architecture, which combines all elements and ICT components in a layer and life cycle model. In RAMI 4.0, the bottom of the six vertical lays (as the layered Z- dimension) is deemed as Physical/Virtual Entity of Asset that brings us value on certain occasions.

The X-axis represents producing related stages, e.g., the Type stage and the Instance stage along with the first dimension (X-dimension in Fig.8 ), as the dimension of Product Life Cycle[7].

---

[7] "RAMI4.0 – a reference framework for digitalization", where Product Life Cycle Value Stream is introduced as the Standardisation "IEC62890".

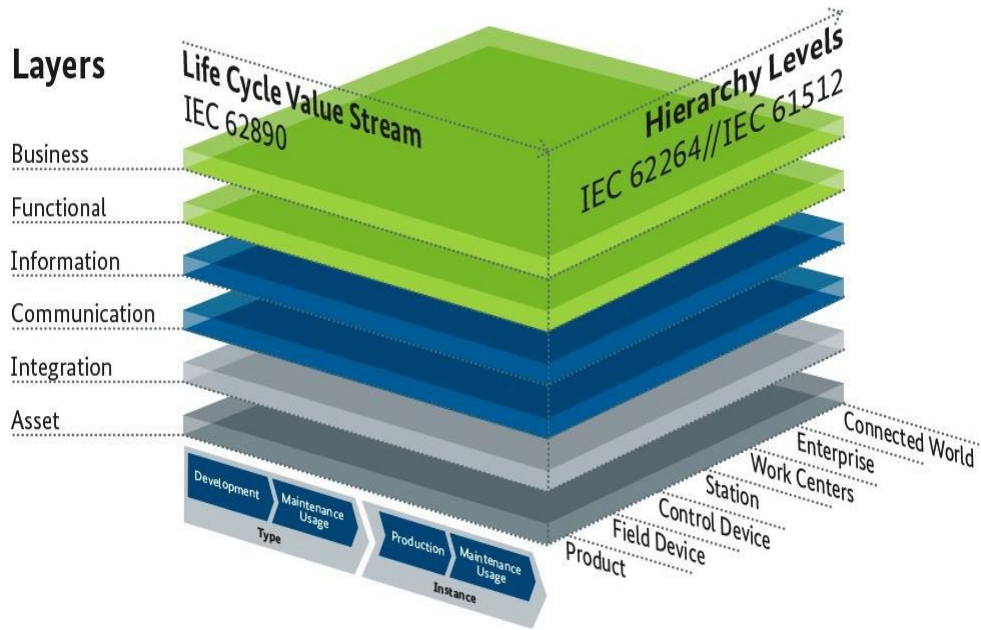

Figure 8　RAMI4.0, where show the attribution of Connecting as a service in y-axis of hierarchy level. (From *Anna Salari*)

The layered architecture axis represents different views of the product or service, such as the business view, the functional view, the communication view and the asset view. The production lifecycle axis covers the full lifecycle of a product/service while taking all participants such as supplier and integrator into account.

RAMI 4.0 provides a standard framework for tracking details of a product or a service, and it breaks down complex processes into easy-to-grasp packages. For example, a safety function (including data privacy and IT security) in the hierarchy axis can be mapped to relevant devices. In the architecture axis, the function is described in detail in the functional layer. In order to achieve the required safety goal, different production phases may be involved, e.g. the material quality needs to be defined in the design phase while being guaranteed in the purchase/ delivery phase.)

In the complex yet needed 3D-model as RAMI4.0, along with the production lifecycle axis (X-axis) as decomposed in Figure 9, Primitive IoTs are "assembled"(e.g. computer simulation) before the phrase "Type" stage, and after their integration/unifying into IoT grids, a "Type" is prepared to be upgraded to an "Instance" which includes a sell-ready production and the maintenance after sell.

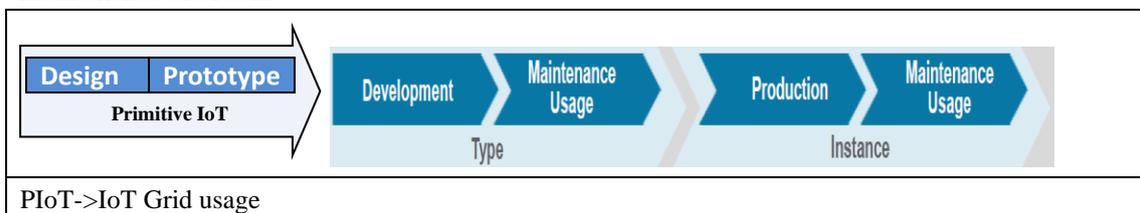

PIoT->IoT Grid usage

Figure 9 the preparing stage as IoT Grid before "Type" stage, when "**Product/service as Primitive IoT**" is designed

The preparing stage plays the role of fundamentals of each layer (e.g. assets/integrations in perception layers) of IoT and each stage of the production lifecycle in RAMI4.0. From the information view, even the virtual "asset" offer value (e.g. not only the patent or idea, but also the program or app./service) and it also need preparation for the design of PIoT and its logical "location" (i.e. where the requirement of a service happens) at a certain IoT Grid before its construction plan.

The compromising target (of IoT Grids) may not be the described product or service directly but only be a feature of it (e.g. intelligence/security/availability/integrity or just an involved control device in the hierarchy axis). For example, a comprehensive attack can be achieved within the whole scope of the RAMI model. It may start from a lower hierarchy level to compromise a higher-level target (e.g. a factory station). In the architecture axis, it comes with its own business purposes, compromises important functions as well as goes through a specific communication topology. At last, the attack can be deployed in the beginning of the lifecycle and will be only triggered at a given point, as part of an Advanced Persistent Threat (APT)[5].

So the Security IoT Grids may be deployed one by one along with the hierarchy axis/ the architecture layer (e.g. communication layer). Accordingly, the security analysis should not be limited within a single Security IoT Grid of a certain level, or layer of phase (or stage of life cycle) of the RAMI4.0 model, as well as IMSA (Intelligent Manufacturing System Architecture of China, see Figure 10.) model.

Either RAMI4.0 or IMSA aims at Common Standards Requirements for digitized industrial production with hierarchy axis aspects----Common connecting by upgrading level, either hierarchy level of RAMI4.0 or system hierarchy of IMSA:

物联网网格（IoT Grids）的折衷目标可能不是直接描述的产品或服务，而只是其一个特征（例如可用性/完整性或只是层次轴中的一个相关控制设备）。其次，可以在RAMI模型的整个范围内实现综合攻击。它可能从较低的层次结构开始，以损害更高层次的目标（例如工厂工作站）。y-axis 不变，产业与应用维度沿产品全生命周期和物理实体的应用解耦为平面

The integration/connecting from product/field device to control device and to distributed control in station/work center, then the sharing value spread from distributed work centers to Enterprise and to the connected world, where they are regulated by IEC62264/IEC61512, including network and protocols; common rules for cyber security and data protection; □Common language (including signs, alphabet, vocabulary, syntax, grammar, semantics, pragmatics and even culture) and so on.

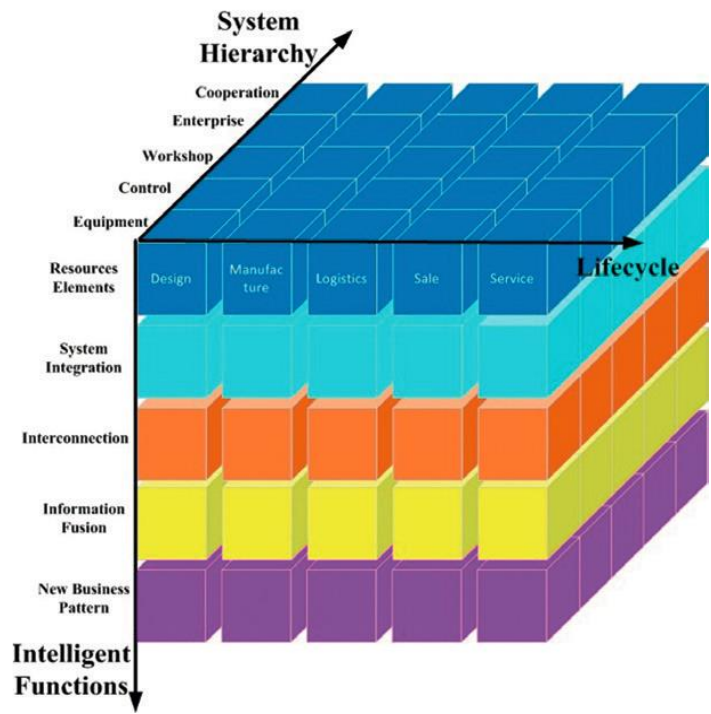

Figure 10 Intelligent Manufacturing System Architecture of China (From: Sino-German White Paper on Functional Safety for Industry 4.0 and Intelligent Manufacturing[8])

The Intelligent Function axis of IMSA (Figure Y-1), as contrast to the Layer axis of RAMI4.0, not only considers technological assets (resource element) and platforms in the view of layered architecture, but Integration/Interconnecting/Fusion of information in Security Networks, CPSs, IoT Grids, Security Cloud/Edge/Big Data and Applications Marketplaces are considered. Take security as an example, in dealing with an unexpected attack, those IoT Grids means the search space of checking point is significantly increased. However, for the given attack, the RAMI or IMSA model with IoT Grids insight provides a break-down view to analyze, discover and prevent or mitigate the attack with the knowledge of the composed Grids inside.

## 4. Healthy Workforce: TRW Use Case as a Healthy Grid of IIoT

With the emergence of a European aging workforce [8], active and healthy protection is one of the major societal concerns in the smart home/city and smart industry scenarios. In fact, human and organizational factors are still involved in almost 90% of all workplace accidents and incidents.

Aging workforce may exhibit more limited physical and cognitive responses. According to the World Health Organization, European workforce will be older than ever before, it will make up for 30% of the working-age population[9].

The European Commission estimates that Muscular-Skeletal Disorders (MSD) accounts for 50% of all absences from work lasting 3 days or longer and for 60% of permanent work incapacity. Furthermore, up to 80% of the adult population will be affected by an MSD at some time in their life. In the last few years, it has become apparent that this situation will even worsen .For this reason, it becomes necessary to improve the quality and comfort of the workforce .In order to develop new solutions and take benefit from the IoT technologies.

---

[8]"https://www.plattform-i40.de/PI40/Redaktion/EN/Downloads/Publikation/industrie-4-0-sino-german-white-paper-on-functional-safety-for-industry-4-0-and-intelligent-manufacturing.pdf?__blob=publicationFile&v=2"

TRW (pronounced alike Tier 1 – automotive supplier, in German) trial aims to develop an IoT-based worker-centric safety management systems, whose aim is to reduce accidents and incidents in the production workplace.

The TRW as a use case of IIoT application is focused on the monitoring and assessment of the ergonomic risk that can affect to the blue-collar workers on the production lines, in order to perform more effective prevention strategies than the traditional prevention strategies.

Current procedures and systems are not customized to the limitations or characteristics of the workers, so the results are not trustworthy when customizing specific plans and current human-based surveillance are not effective enough. So the TRW trial specially provides the following functionalities: i) real-time data collection through Kinect sensors (see Figure 11a), ii) continuous data processing for ergonomic risks detection, iii) events management, and iv) web services and applications for corrective actions performance [3](e.g. web services for sending messages or warnings related to the risks detected; see Figure 11b).

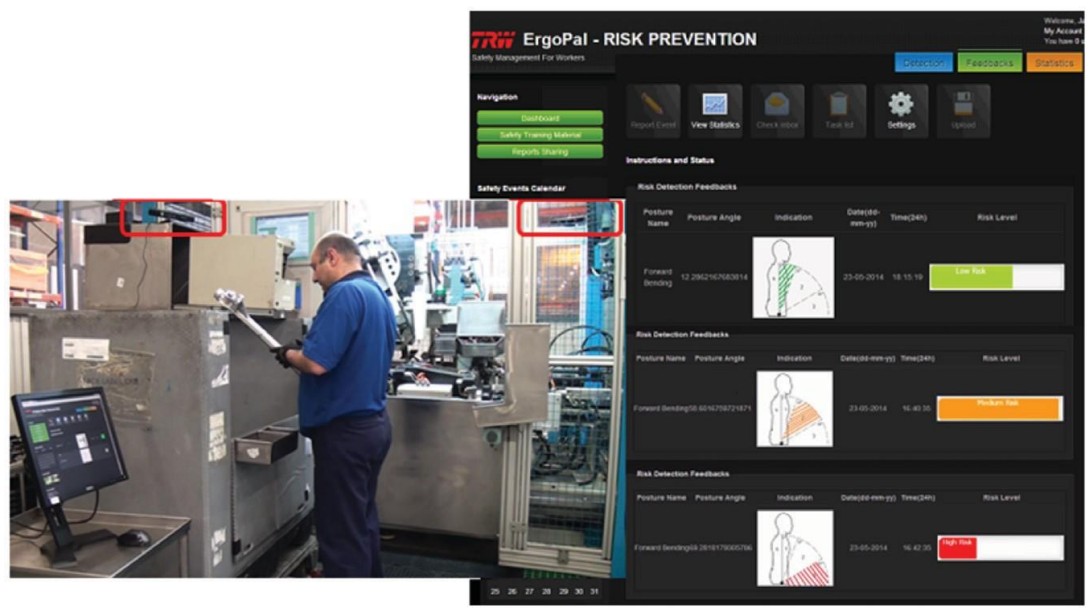

**Figure 11**   TRW trial real environment (a) and GUI (b).

The platform of TRW Trial (see Figure 12) is based on the Smart Factory Platform (GEs: General Ends, and SEs: Special Ends), and new components and services developed for the trial TRW Platform, which takes advantage of functionalities related to data gathering, complex event processing, context information management, and event information delivery, among components related to IoT service, e.g. monitor, sensors data repository, risk/action repository, etc. in Figure 12.

System monitors and collects real-time data of the skeleton and movements of the workers in the production line through Kinect sensors. **Risk/action repository service collects and** processes the data in order to detect "unsafety" events based on the defined rules (e.g. angles of the body or frequency of movements higher than the threshold values). In case of event detection, the different services or actions previously set up are delivered, e.g. messages or warnings are sent to the blue-collar workers and prevention technicians.

Preliminary intermediate KPIs already demonstrate a reduction of 13% in the number of accidents and incidents in the factory, as well as a performance improvement of 80% in the number of risks detected and alarms activated[3].

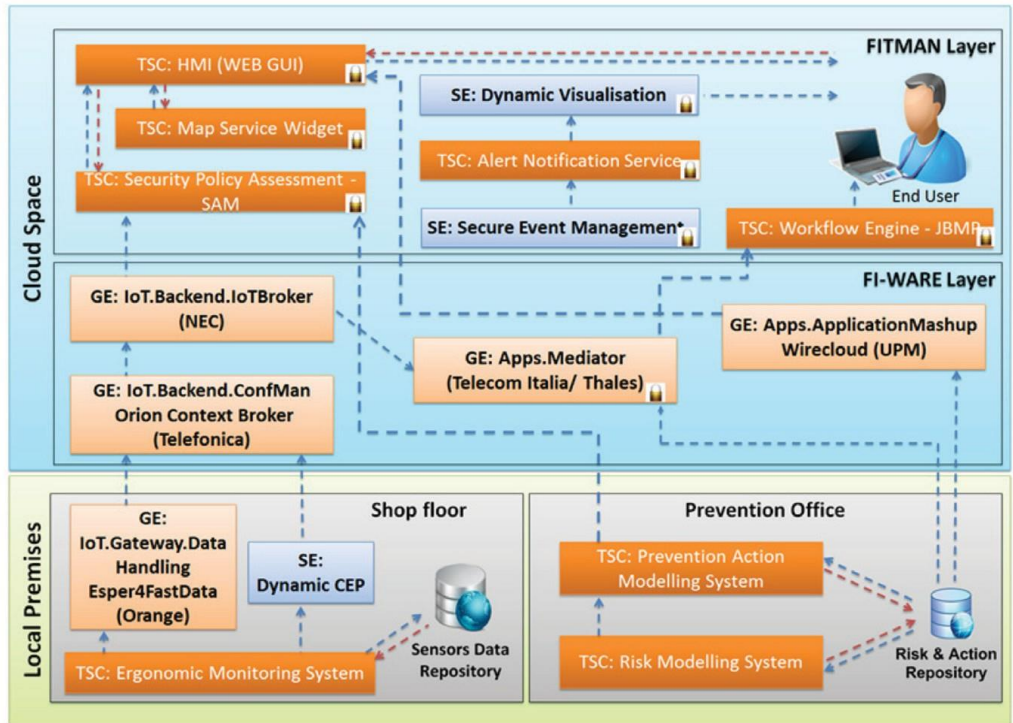

**Figure 12** Future Internet Platform for Safer and Healthier Workplace[9].

The 3rd service interface in the cloud can be seen as map service widget and security policy assessment in Fig.12. TRW trial introduces and accepts IoT technologies in the local manufacturing in presence，as well as the cloud (e.g. workflow management, alert notification service, secure event management, etc.), in order to get the workplace of the near future of IoE era, providing the necessary balance between security concerns and privacy concerns. The implementation of IoT technologies of industry supports the innovative human-in-the-loop model, getting a participative approach for the workers' empowerment solution, and improving the value chain of IIoT.

# 5. Conclusion and Discussion

The emergence of the IoT with its billions of envisioned devices poses a clear challenge to the design, management, production and maintaining of these, which are expected to extended to the vertical value chains as different IoT industries (or business) and to the horizontal value chains as layered platform (for interaction), where are from perception/ assets, networks/communication, software/data platform, to digital service/application.

IIoT covers the emerging approach in industrial environments, which is to create system intelligence by a large population of intelligent, small, networked, embedded devices at a high level of granularity, as opposed to the traditional approach of focusing intelligence on a few large and monolithic applications within industrial solutions.

IoT industries are broader than Industrial IoTs covers. But Industrial IoTs are more complex from the view of IoT RAMEC (Reference Architecture Model for Edge Computing, presented at the 3rd

---

[9] Ovidiu Vermesan, Peter Friess. Building the Hyperconnected Society IoT Research and Innovation Value Chains, Ecosystems and Markets[M]. River Publishers.2019 P159

Edge Computing Forum, by the European Edge Computing Consortium) and IIC (Industrial Internet Consortium) reference models and architectures. One reason of this part is that the fragments of most IoT required to be united in certain organizing view, i.e., Primitive IoT presented in the reference [1] is a minimized service set, which construct/compose the IoT Grid Layer as components of IoT Zone. Another reason is that the organized view from IoT Grid to IoT Zone in the IoT-A[9] and IIC references architectures is reused similarly, in the bottom-up view from Primitive IoT to IoT Grid. So concepts from Primitive IoT to IoT Grid and to IoT Zone are described to be better understood in a unified view (architecture views, viewpoints, stakeholders, concerns, etc.) Along with the upgrading level, connection as a service in IIoT may be supported by 5G slice , such as Edge computing for IoT grid, from devices/equipment/ things on site, to workshop/enterprise edge, and to the cloud .cooperation.

At last, this paper gives an healthy IoT case as an example, where the Intelligent Function(in Figure 10) is included as Risk/action repository service, which collects and processes the data in order to detect "unsafety" events. From the Insight of Artificial Intelligence(such as Machine Learning), forecasting is a key issue in the prominent IoT use case of predictive maintenance, as well as the predictive for human diagnose, which is used to determine the health of a piece of machine (equipment/asset) and understand when any maintenance/exercise might be needed. Forecasting involves predicting new outcomes based on previously known results. Depending on the IoT use case, different forecasting timeframes apply. For example, trajectory forecasting of moving objects can be real time, whereas machine degradation is more long term. Forecasting can be data driven or model driven depending on the problem requirements. **Model training is a necessity in IIoT application and can be based on training sets or via reinforcement learning. Typical forecasting models can be statistical or neural networks-based, Bayesian or non-Bayesian, linear or nonlinear, parametric or nonparametric, univariate or multivariate.**

# Acknowledgments

The content of this paper is largely relying on the contributions of the IoT workgroup of National University of Defense Technology. The authors would like in particular to thank the funds of China Post-doctoral as No.sz32343.

1*Hang song, he is

2 YUNCHENG JIANG received the B.S. degree in automotive engineering from Southwest Jiaotong University, Chengdu, China, in 2016. He received the first M.S. degree in mechanical engineering from Clemson University, Greenville, USA, in 2017, and second in automation from Hongkong University of Science and Technology, Hongkong, China, in 2019. He is currently an advanced engineer with Harbor Technology, Ltd.
    His main research interests include planning, control, numerical optimization and convex optimization.